\documentclass[a4paper, 12 pt]{article}
\usepackage{amsmath}
\usepackage{amsfonts}
\usepackage{amsthm}
\usepackage{amscd}
\usepackage{amssymb}
\usepackage{color}
\usepackage{graphics}

\newcommand{\half}{{\textstyle \frac{1}{2}}}
\newcommand{\fourth}{{\textstyle \frac{1}{4}}}

\newcommand\smallfrac[2]{{\textstyle \frac{#1}{#2}}}

\newcommand{\phivec}{{\boldsymbol \phi}}
\newcommand{\phitilde}{{\tilde \phi}}
\newcommand{\phivectilde}{{\tilde{\boldsymbol \phi}}}
\newcommand{\cvec}{{\bf c}}

\newcommand{\evec}{{\bf e}}
\newcommand{\nvec}{{\bf n}}
\newcommand{\pvec}{{\bf p}}
\newcommand{\uvec}{{\bf u}}
\newcommand{\npvec}{{\bf n'}}

\newcommand{\nuvec}{{\boldsymbol \nu}}

\newcommand{\svec}{{\bf s}}

\newcommand\sgn{\,{\hbox{\rm sgn}}}

\newcommand{\avec}{{\bf a}}
\newcommand{\bvec}{{\bf b}}

\newcommand{\rvec}{{\bf r}}
\newcommand{\rvecp}{{\bf r'}}
\newcommand{\vvec}{{\bf v}}
\newcommand{\wvec}{{\bf w}}
\newcommand{\xvec}{{\bf x}}

\newcommand{\Hvec}{{\bf H}}
\newcommand{\Fvec}{{\bf  F}}

\newcommand{\Rr}{{\mathbb R}}

\newcommand{\czero}{{ C_{\mbox{\gotic T}}^0}}
\newcommand{\ctwo}{{ C_{\mbox{\gotic T}}^2}}
\newcommand{\ck}{{ C_{\mbox{\gotic T}}^k}}
\newcommand{\cinf}{{ C_{\mbox{\gotic T}}^\infty}}
\newcommand{\Lip}{\text{Lip}}

\newcommand{\inv}{\,\text{\rm inv}}

\newcommand\Ltilde{{\tilde L}}
\newcommand\nvectilde{{\bf \tilde n}}

\newtheorem{thm}{Theorem}[section]
\newtheorem{lem}{Lemma}[section]

\theoremstyle{definition}
\newtheorem{defn}{Definition}[section]

\theoremstyle{remark}
\newtheorem{rem}{Remark}[section]

\newcommand{ \dst} {\displaystyle}

\font\gotic=eufm10 scaled 700
\newcommand{\lt} {\ {\bf <}}
\newcommand{\rt} {{\bf >}\ }

\newcommand{\lb} {\left(}
\newcommand{\rb} {\right)}

\newcommand{\lbr} {\left\{}
\newcommand{\rbr} {\right\}}
\newcommand{\ls} {\left[}
\newcommand{\rs} {\right]}

\newcommand{\ld} {\left.}
\newcommand{\rd} {\right.}
\newcommand{\lv} {\left|}
\newcommand{\rv} {\right|}

\newcommand{\al} {\alpha}

\newcommand{\ga} {\gamma}
\newcommand{\bga}{\begin{array}{l}}
\newcommand{\ena}{\end{array}}
\newcommand{\bge}{\begin{equation}}
\newcommand{\ene}{\end{equation}}

\def\comment#1{}

\def\withcomments{
\addtolength{\oddsidemargin}{-0.5 in}
\addtolength{\evensidemargin}{-0.5 in}
\newcounter{mycommentcounter}
\def\comment##1{\refstepcounter{mycommentcounter}%
  \ifhmode%
  \unskip%
  {\dimen1=\baselineskip \divide\dimen1 by 2 %
    \raise\dimen1\llap{\tiny -\themycommentcounter-}}\fi%
  \marginpar{\renewcommand{\baselinestretch}{0.8}%
    \footnotesize [\themycommentcounter]: \raggedright ##1}}
}

\withcomments

\begin{document}
\date{8 June, 2004.}
\title{Lower bound for energies of harmonic tangent unit-vector fields on convex
polyhedra}
\author{A Majumdar$^{\dag,\ddag}$, 
JM Robbins$^{\dag}$ \& M Zyskin$^{\dag}$\thanks{  {\tt
  m.zyskin@bristol.ac.uk}}\\
$^\dag$School of Mathematics\\ University of Bristol, University Walk, 
Bristol BS8 1TW, UK\\
$^\ddag$  Hewlett-Packard Laboratories\\
 Filton Road, Stoke Gifford, Bristol BS12 6QZ, UK}
\maketitle
 
%\begin{abstract}
\noindent  {\bf Abstract.} We derive a lower bound for energies of harmonic maps of convex
  polyhedra in $ \Rr^3 $ to the unit sphere $S^2,$ with tangent
  boundary conditions on the faces.  We also establish that $C^\infty$
  maps, satisfying tangent boundary conditions, are dense with respect
  to the Sobolev norm, in the space of continuous tangent maps of
  finite energy.
  
\noindent  {\bf Mathematics Subject Classifications (2000).} 58E20, 35J55.

\noindent  {\bf Key Words.} Bi-stable nematic devices, liquid crystals, harmonic maps, polyhedra, tangent boundary conditions, lower bounds.
%\end{abstract}

\newpage
%\vspace{-2cm}
\tiny
\hspace*{70mm} \mbox{"The unburdened mind is joy supreme"}
\\
\hspace*{80mm} \mbox{Gel-sang-gya-tso, 1708-1757.}
\normalsize

\section{Introduction}\label{sec: Intro}

We consider maps $\nvec$ from a convex polyhedron $P \subset \Rr^3 $
to the unit sphere $S^2$, which we regard as unit-vector fields on
$P$.  $\nvec$ is said to satisfy {\em tangent boundary conditions}, or
to be {\em tangent}, if, on the faces of $P$, $\nvec$ is tangent to
the faces.  This implies that, on the edges of $P$, $\nvec$ is
parallel to the edges, and, therefore, discontinuous at the vertices.
We consider maps which are continuous away from vertices and which
belong to the Sobolev space $W^{1,2}(P)$. There are homotopically
inequivalent classes of such maps \cite{rz}.  In this paper, we derive
a lower bound for the energy 
\begin{equation}\label{eq: one-constant}
E[\nvec] = \int\int\int_P \lb \nabla \nvec\rb^2 dV
\end{equation}
for each homotopy class.  We also establish that $C^\infty$ tangent
unit-vector fields are dense in the space of continuous tangent
unit-vector fields in $W^{1,2}(P)$ with respect to the Sobolev norm.

This work is part of a study of liquid crystals in polyhedral
geometries started in \cite{rz}.  We have been motivated by
applications to the design of bi-stable liquid crystal displays (see,
eg, \cite{newtonspiller}) as well as by mathematical considerations.

A nematic liquid crystal is a suspension
of rod-shaped molecules in a liquid substrate.  The molecules have a
preferred average orientation at every point in space.  This preferred
orientation is described by a {\it director field} -- a unit-vector
field with opposite orientations identified  %The director field plays
%the role of an order parameter, in the spirit of the Landau-Ginzburg
%theory of phase transitions 
\cite{dg}.  We are
only considering continuous director fields in a simply connected
domain, in which case an orientation can be chosen arbitrarily at one
point and defined elsewhere by continuity, thus yielding a unit-vector
field $\nvec$.  (Quasi-) stable configurations are (local) minima of a
certain energy functional, the Frank energy \cite{dg}
\begin{align}
  \label{eq:Frank}
  E_F[\nvec] = \int\int\int_V &\big( K_1 (\text{div}\, \nvec)^2 + K_2 (\nvec \cdot
  \text{curl}\, \nvec)^2 + K_3 (\nvec \times \text{curl}\, \nvec)^2 +\notag\\
&\quad   + K_4 \,\text{div}\,  \ls ({\mathbf{n}}\cdot \nabla) {\mathbf{n}} -
{\mathbf{n}}\, \text{div}\, {\mathbf{n}} \rs \big)
  \,dV.
\end{align}
In the  `one-constant approximation' $K_1=K_2=K_3=K_4=1$, and the Frank energy
reduces to \eqref{eq: one-constant}.

Solutions of the Euler-Lagrange equation corresponding to \eqref{eq:
  one-constant}, subject to appropriate boundary conditions, are
harmonic maps of $P$ to $S^2$.  The boundary conditions are determined
by the substrate and surface treatment used.  In the cases being
considered here, it is strongly energetically favorable for the vector
field to be tangent to the boundary.

%For polyhedral
%domains, this implies that each face of $P$ is mapped to the great
%circle of $S^2$ parallel to the face.  Such tangent unit-vector fields
%are necessarily discontinuous at the vertices; we assume continuity
%everywhere away from the vertices. 

Let $\bar{P} \subset \Rr^3$ be a convex
%compact
polyhedron.  
Let $\vvec^a$, $a = 1,
\ldots, v$, label the vertices, $E^b$, $b = 1,\ldots, e$, the edges
and $F^c$, $c = 1,\ldots, f$, the faces of $P$ respectively. $ \Fvec^c$
is the outward unit-normal vector to $F^c.$ Let $P $ denote
$\bar{P}$ without its vertices.  
%We identify unit-vector fields
%$\nvec: P\rightarrow \Rr^3, \vert \nvec \vert =1 $ with maps to the
%unit sphere $S^2\subset \Rr^3.$ 
%Let 
%$\mbox{\goth T}$ of tangent unit-vector fields on $P.$ 
Let $\czero(P)
$ denote the space of continuous tangent unit-vector fields on $P$.
We shall also have occasion to refer to
$\ck(P)$, the space of $C^k$ tangent unit-vector fields, and
$\cinf(P)$, the space of smooth tangent unit-vector fields.
%equipped with the compact-open topology.  %JMR

We say that   $\nvec$, $\npvec \in
\czero(P)$ are  homotopic,  $\nvec \sim \npvec$, if there exists
a continuous map $\Hvec: P\times [0,1]\rightarrow S^2;
(\xvec,t)\mapsto \Hvec_t(\xvec)$, such that $\Hvec_t \in\czero(P)$ for
$t \in [0,1]$
   and
$\Hvec_0 = \nvec$, $\Hvec_1 = \npvec$.
%\newpage

It is shown in \cite{rz} that homotopy classes of tangent unit-vector 
fields are classified by a set of invariants, which we call 
{\it edge orientations}, {\it kink numbers} and {\it wrapping
  numbers}.  These are defined as follows:\\
\noindent {\em Edge orientations:} The edge orientation $\evec^b(\nvec)$ is the
value of $\nvec$ on the edge $E^b.$

\vspace{1mm}
\noindent {\em Kink numbers:} 
%Each face $F^c$ is mapped to the great circle
%$C^c\subset S^2$ which is parallel to $F^c$.
Let $\ga^{ac}$ denote a path on $F^c$, positively oriented with
respect to $\Fvec^c$, between the pair of edges with common
vertex $\vvec^a$.  The image of $\ga^{ac}$ under $\nvec$ describes an
arc on $C^c$, the great circle in $S^2$ parallel to $F^c$.  The kink
number $k^{ac}(\nvec)$ is the degree of the closed path on $C^c$
obtained by closing $\nvec(\ga^{ac})$ with the shortest arc between
its endpoints.  Since $\nvec$ is continuous away from vertices, its restriction
to any closed path on $F^c$ away from vertices has degree
zero.  This implies the following sum rule 
for the kink numbers: Let
$q^c(\nvec)$ denote the number of vertices of $F^c$ at which the edge
orientations are oppositely oriented with respect to the normal
$\Fvec^c$.  Then $\sum_{\vvec^a \in F^c} k^{ac}(\nvec) = 1- \half q^c(\nvec)$.
%% \bge
%%   \label{eq:kink sum rule}
%%   \sum_{\vvec^a \in F^c}
%% k^{ac}(\nvec) = 1- \half q^c(\nvec).
%% \ene
\vspace{1mm}

\noindent {\em Wrapping numbers.} Choose $\svec\in S^2$ such that
$\svec$ is not tangent to any of the faces of $P$.
% JMR. don't need to include -s
For each vertex $\vvec^a$, choose an outward-oriented surface
$S^a\subset P$ which separates $\vvec^a$ from the other vertices.  The
boundary of $S^a$ lies on those faces of $P$ which meet at $\vvec^a$.
%$w_a$ can be parameterized by a
%unit disk.
%, so that boundary of $w_a$ is parameterized by boundary of the disk.
Construct a new map $\nuvec^a: S^a \rightarrow S^2$ which coincides
with $\nvec$ on $\partial S^a$ and whose image does not contain $\svec$.
%
%. on points of $w_a$ parameterized  by a
%radial ray of the disk describe the shortest geodesic  on the target sphere,
%connecting image of the corresponding boundary point and the point
%$-s$. 
The wrapping number $w^a(\nvec)$ is the degree of the map
$S^2\rightarrow S^2$ obtained by gluing the maps $\nvec|_{S^a}$ and
$\nuvec^a$ along the boundary of $S^a$.  
The fact that $\nvec$ is continuous on 
$P$ implies that $\sum_{a=1}^v w^a(\nvec) = 0$.
%\begin{equation} 
%\label{eq: wrap sum}
%\sum_{a=1}^v w^a(\nvec) = 0.
%\end{equation}
%\end{prop}
%The edge orientations, kink numbers and
%wrapping numbers are homotopy invariants and, subject to the sum
%rules (\ref{eq:kink sum rule}) and (\ref{eq: wrap sum}), they classify
%$\czero(P)$ up to homotopy.  

In what follows, we denote the invariants collectively by $\inv =
\{\evec^b, k^{ac}, w^a\}$.  

The paper is organized as follows.
A lower bound for the energies of
$C^\infty$ tangent unit-vector fields in terms of the invariants is
given in Theorem~\ref{th: th1}.  The derivation adapts methods of
\cite{bl} to the tangent boundary-value problem treated here.
(Boundary value problems were not considered in \cite{bl}. Also, the
lower bound involves not only the degree of the map on two-dimensional
surfaces surrounding the vertices, but also kink numbers and edge
orientations).  In Theorem~\ref{thm: thm 2} we establish that
$C^\infty $ tangent unit-vector fields are dense in the space of
continuous Sobolev tangent unit-vector fields with respect to the
Sobolev norm.  The proof requires certain smoothings of the vector
fields which lie outside the scope of the standard Meyers-Serrin
theorem \cite{adams}.  Thus, the
lower bound of Theorem~\ref{th: th1} extends to $\czero(P)\cap
W^{1,2}(P)$.  
Section~\ref{sec:discussion} contains a discussion of the results.

\section{Lower bounds for energies of harmonic maps}\label{sec: energies}
\begin{defn}\label{def:energy}
  The minimal energy $M(h)$ of maps in homotopy class $h$ is defined by
\begin{equation} 
M(h) = \inf_{ \genfrac{}{}{0pt}{}{\nvec \in \czero (P) \bigcap
      W^{1,2} (P),}{ \text{ inv} (\nvec) = h}} E [ \nvec],
\label{energyMinimimal}
\end{equation}
where $W^{1,2}(P)$ is the Sobolev space
\begin{equation}\label{eq: Sobolev}
W^{1,2}(P)= \lbr \nvec\left\vert \rd 
{\nabla }\nvec \in L^2(P) \rbr 
\end{equation}
and
\bge
E[\nvec] =  \int\int\int_P (\nabla \nvec)^2\,dV= 
\dst\int\dst\int\dst\int_P  \partial_a n_b \partial_a n_b
\,dV=
\|
\nvec\|^2_{W^{1,2}(P)}.
\label{energy}
\ene
\end{defn}

%\noindent {\bf Remark.} 

If the infimum is actually achieved by some tangent unit-vector
field $\nvec \in \ctwo(P), $ then $\nvec$ satisfies the Euler-Lagrange
equation \bge \Delta \nvec - \lt \nvec, \Delta \nvec \rt \nvec =0,
\ene with boundary conditions $\ld \nvec \cdot {\Fvec}^c \rv_{F^c}
=0,$ $((\Fvec^c\cdot\ld \nabla)\nvec) \cdot \lb \nvec \times
\Fvec^c\rb\rv_{F^c} =0$ on the faces of $P.$
%
%Below we obtain a lower bound for $M(h)$.  The derivation adapts
%\cite{bl} to our tangent boundary-value problem.  
It will be shown in Section~\ref{sec: approximation} that $\cinf
(P)\bigcap W^{1,2} (P) $ is dense in $\czero (P)
\bigcap W^{1,2} (P)$ with respect to the Sobolev norm.  Thus, to
compute $M(h)$, it suffices to consider smooth maps only.

It is straightforward to show (a demonstration is given in \cite{bl}) that 
the energy density $\rho = (\nabla \nvec)^2$ satisfies the inequality 
%It is straightforward to show that the energy density $\rho = (\nabla
%\nvec)^2$ satisfies the inequality 
\begin{equation}
%\bga
 \rho \geq
2 \left|\nvec^* \omega \right|,
\label{e>d}
%\ena 
\end{equation}
where $\omega$ is the area-form on $S^2$, normalized to have
area $4\pi$, and $|\nvec^*\omega|$ is the Euclidean norm of its
pull-back.  Indeed, since $(\nvec\cdot\nabla)\nvec = 0$, 
we may write
\begin{equation}\label{eq: alphas}
 \nabla
\nvec = \alpha_1 {\boldsymbol \tau}_1 + \alpha_2 {\boldsymbol \tau}_2,
\end{equation}
where $ {\boldsymbol\tau}_1$, ${\boldsymbol \tau}_2$ constitute a
(locally defined) orthonormal basis for $T_{\nvec}S^2$, and
$\alpha_1$, $\alpha_2$ are (locally defined) one-forms on $P$. It
follows that $\rho = |\al_1|^2 + |\al_2|^2$, while $|\nvec^* \omega |
= \left|\al_1\wedge \al_2\right|^2 = |\al_1|^2 |\al_2|^2
-(\al_1\cdot\al_2)^2$, where $\al_1\cdot\al_2$ denotes the Euclidean
inner-product on forms.  Therefore, $\rho^2- 4 \left| \nvec^* \omega
\right|^2 = (|\al_1|^2 -|\al_2|^2)^2 + 4 (\al_1\cdot\al_2)^2 \geq 0.$

\begin{rem} 
Suppose $\alpha_1\wedge\alpha_2 \ne 0$. 
From (\ref{eq: alphas}), if $\iota_X (\alpha_1\wedge\alpha_2) = 0$
for some vector field $X$, 
then 
$\iota_X \nabla \nvec = 0$; ie, $\nvec$ is constant on
the characteristics of $\al_1\wedge \al_2.$
\end{rem}
\begin{rem}
  If we have equality in (\ref{e>d}), then $|\al_1|^2 =|\al_2|^2$ and
  $\al_1\cdot\al_2=0,$ so that $\nabla \nvec$, regarded as a map from
  the orthgonal complement of the characteristic distribution of
  $\al_1\wedge \al_2$ to $T_\nvec S^2$, is conformal.
\end{rem}

\begin{defn}
The {\it trapped area}  $\Omega^a(\nvec)$ at a vertex $\vvec^a$ is the
area
(as a proportion of the area of $S^2$)
 of the image under $\nvec$ of an outward-oriented surface, $S^a \subset P$,  which
 separates $\vvec^a$ from the other vertices, ie
\begin{equation}\label{eq: Q}
\Omega^a(\nvec) = \frac{1}{4\pi} \int\int_{S^a} \nvec^* \omega.
\end{equation}
\end{defn}

The trapped areas are homotopy
invariants, and may be expressed in terms of edge orientations, kink
numbers and wrapping numbers as follows (see \cite{rz} for details).
Given a vertex $\vvec^a$, let $K^a$ be the geodesic polygon on $S^2$ with
vertices $\evec^{b_1},\ldots \evec^{b_{m}}$ given by the edge
orientations of the edges $E^{b_1},\ldots, E^{b_m}$ which are incident at
$\vvec^a$ (the edges are ordered consecutively with respect to the
outward normal on $S^a$).  Then
\begin{equation}
\Omega^a =
 w^a  -  \half {\sum_{c}}'
\sgn(\Fvec^c\cdot\svec) k^{ac}
+  \sum_{j=2}^{m-1} \left(\frac{1}{4\pi}A(\evec^{b_1},\evec^{b_j},\evec^{b_{j+1}}) - 
\sigma(\evec^{b_1},\evec^{b_j},\evec^{b_{j+1}})\right),
\label{Omega[invariants]}
\end{equation}
where the sum $\sum_{c}'$ is taken over the faces $F^c$ incident at
$\vvec^a $, $A(\avec,\bvec,\cvec)\in (-2\pi,2\pi)$ is the oriented
area of the spherical triangle with vertices $\{\avec$, $\bvec$,
$\cvec$\} and $\sigma(\avec,\bvec,\cvec)$ is equal to $\sgn
((\avec\times\bvec)\cdot \svec)$ if $\svec$  is
contained in the spherical triangle with vertices $\{\avec$, $\bvec$,
$\cvec$\} and is zero otherwise.
The trapped areas are typically not integer-valued.  However, they
%satisfy the sum rule, analogous to (\ref{eq: wrap sum}),
satisfy the sum rule (related to the sum rule for the wrapping numbers)
\begin{equation}\label{eq: trapped sum}
 \dst\sum_a \Omega^a   = 0
\end{equation}
(this follows from the fact that the map $\nvec:\partial P\rightarrow
S^2$  is contractible).

%JMR - removed this
%Since for   surfaces $s_{v_a}$   as above,
% \bge
% \int\int_{s_{v_a}} \rho \geq 2   \int\int_{s_{v_a}}  \left| \nvec^* \omega
%\right| \geq \vert  8\pi  \Omega_i \vert,
% \ene
%it is clear that the minimal energy   is bounded from below  by a positive
%constant.
%\\

\begin{thm}
  \label{th: th1}
The minimal energy $M(h)$ 
is bounded below  by  
\begin{equation}
 M(h) \geq \max_{\lbr \xi^a:  \vert \xi^a-\xi^{a'}\vert  \leq 
\left|\vvec^a - \vvec^{a'}
\right|
\rbr} 
\lb8\pi \dst\sum_a \Omega^a \xi^a\rb > 0.
\label{low_bound}
\end{equation}
\end{thm}
\begin{rem}
The bound is given in terms of a finite-dimensional linear
optimization problem with linear constraints, whose solution can be
found algorithmically using standard methods.
\end{rem}

\begin{proof}
  Let $\nvec \in \cinf(P)\cap W^{1,2}(P)$ with $\inv(\nvec) = h$.  Let
  $\xi \in \Lip_1(P)$, the space of Lipschitz functions on $P$ with
  Lipschitz constant less then one.  Then $\xi$ is almost-everywhere
  differentiable with $\vert d \xi \vert \leq 1$.  It follows from
  \eqref{e>d}
that
\begin{equation}
E[\nvec] =  \int\int\int_{P} \rho \,dV \geq 2  \int\int\int_{P} \left|\nvec^* \omega \right|
dV \geq  2  \int\int\int_{P}    d \xi  \wedge \nvec^* \omega.   
\label{eq: xi}
\end{equation}
We remove infinitesimal neighbourhoods of the vertices from the domain
of integration and integrate by parts in the last expression.  As $d\,
\nvec^* \omega = \nvec^* d\, \omega =0$, the volume integral vanishes.
Because $\nvec$ is tangent, $\nvec^* \omega$ vanishes on the faces
$F^c$, so that the only contribution to the surface integral is from
the boundaries of the excised infinitesimal neighbourhoods.  On these
boundaries, $\xi$ can be replaced by its values at the vertices.
Recalling (\ref{eq: Q}), we obtain
\begin{equation}
  \label{eq:identity}
   \int\int\int_{P}  d \xi  \wedge \nvec^* \omega = 8\pi \dst\sum_a \Omega^a
\xi(\vvec^a).
\end{equation}
Since (\ref{eq: xi}) and (\ref{eq:identity}) hold for all $\nvec\in
\cinf(P)\cap W^{1,2}(P)$,
it follows that 
 \begin{equation}
 M(h)\geq \sup_{_{\xi \in \Lip_1(P)}}  \lb  8\pi \dst\sum_a \Omega^a
 \xi({
\vvec}^a)
\rb.
\label{xisup}
 \end{equation}

$\xi\in  \Lip_1(P)$ implies the constraints
$\vert \xi(\vvec^a)-\xi(\vvec^{a'})
\vert \leq \vert\vvec^a- \vvec^{a'}\vert$.
Conversely, given a set of 
$\xi^a$'s satisfying these constraints, we can construct a function $\xi \in
\Lip_1(P)$ with $\xi(\vvec^a)=\xi^a$, eg by letting
$\xi(\rvec): = \max_{a} \lb \xi^a -  |\rvec - \vvec^a| \rb$.
Therefore,
\begin{equation}
\sup_{\xi \in \Lip_1} \lb 8\pi \dst\sum_a \Omega^a \xi(\vvec^a)  \rb = 
\max_{\lbr \xi^a:  \vert \xi^a-\xi^{a'}\vert  \leq \vert \vvec^a -
\vvec^{a'}\vert \rbr} 
\lb 8\pi \dst\sum_a \Omega^a \xi^a\rb, 
\label{optimise_xi}
\end{equation}
which together with (\ref{xisup}) gives the required lower bound.
Note that \eqref{eq: trapped sum} implies that
the optimal $\xi^a$'s are determined up to an additive constant, which we
can fix, say, by setting $\xi^1=0.$  The constraints ensure
that the feasible set is nonempty and bounded; 
thus the maximal value in the
right-hand side is finite. It is obvious that that maximal value is
positive, so that 
the lower bound is nontrivial.
\end{proof}
\begin{rem}
  The maximisation problem which appears in the lower bound
  \eqref{low_bound} can be replaced by its equivalent dual
  minimisation problem, as in \cite{bl}.
% the Kantorovich theorem.
\end{rem}

\section{Approximating by smooth tangent maps}\label{sec: approximation}
In Theorem \ref{thm: thm 2} below we show that $\cinf(P)\bigcap
W^{1,2} (P) $ is dense in $ \czero (P) \bigcap W^{1,2} (P)$.  This is
accomplished by constructing, for a given continuous tangent
unit-vector field $\nvectilde$, a smooth tangent unit-vector field
$\nvec$ arbitrarily close to $\nvectilde$ with respect to the Sobolev
norm.  Away from the vertices and edges of $P$, $\nvec$ is obtained
from a smooth average of $\nvectilde$ which preserves tangent boundary
conditions.  Neighbourhoods of the vertices and edges require special
treatment, which is dealt with in Lemmas  \ref{lem: vertex extension}
and  \ref{lem: edge}
respectively.

\begin{lem}\label{lem: vertex extension}
  {\bf Vertex extension.}  Let $\vvec^a$ be a vertex of $P.$ Let
  $l^a$ be a ray from $\vvec^a$ into the interior of $P$, and
  introduce local Euclidean coordinates centred at $\vvec^a$ with
  positive $z$-axis along $l^a$.  Let 
\begin{equation}\label{eq: Lambda^a}
\Lambda^a(H):=\lbr (x,y,z)\in P
  \,\vert\, 0< z \leq H\rbr
\end{equation}
denote the prism obtained by cutting $P$ by a plane perpendicular to
$l^a$ at a distance $H$ from $\vvec^a$.  Let 
\begin{equation}
  \label{eq:Pi}
  \Pi^a = \overline{\Lambda^a(H) \setminus  \Lambda^a(\half H)}
\end{equation}
denote the closed lower half of
$\Lambda^a(H)$.  Given a $C^\infty$ tangent unit-vector field
$\nvectilde$
in $\Pi^a$ and some $\epsilon_3 > 0$ such that 
\begin{equation}
E[\nvectilde]:=\dst\int\dst\int\dst\int_{\Pi^a} \lb \nabla
\nvectilde
\rb^2\, dV \leq\epsilon_3,
\label{n_energyPi_small_vert} 
\end{equation}
one can construct a $C^\infty$ tangent unit-vector field $ \nvec$ on
$\Lambda^a(H)$ coinciding with $\nvectilde$  on \linebreak
$\Lambda^a(H)\setminus\Lambda^a(\smallfrac{3}{4} H)$
such that
\begin{equation}
E[{\nvec}]:= \dst\int\dst\int\dst\int_{\Lambda^a(H)} \lb \nabla
\nvec\rb^2\,dV \leq 
C^a\epsilon_3,
\label{n_energyLa_small_vert}
\end{equation}
where $C^a>0$ is independent of $\nvectilde$, $\epsilon_3$ and $H$.
\end{lem}
\begin{proof}
Let $D := \lbr (u,v) \,\vert \, (Hu,Hv,H) \in P\rbr$ denote the base
of $\Lambda^a(H)$, parameterized by $u$ and $v$.
Introduce new coordinates $(u,v,h)$ by
\bge
\bga
z=Hh, \quad x= Hh u, \quad y =  Hh v ,\quad \text {where}\ 0< h \leq 1, \ \text{and}\ (u,v) \in \mathit{D}.
\ena
\ene
Let $\phivectilde(u,v,h)= \nvectilde( Hh u, Hh v,  Hh).$
From  (\ref{n_energyPi_small_vert}),
\bge
\bga
E[\nvectilde] 
%\\
=H \dst\int_{\frac{1}{2}}^1  dh\dst\int\dst\int_{D}dudv  \lbr  \lb 
\phivectilde_u \rb^2 + \lb  \phivectilde_v \rb^2 + \lb  -u
\phivectilde_u -v \phivectilde_v + h
\phivectilde_h\rb^2
\rbr  \leq \epsilon_3.
\label{n_energyPi_newcoor_vert}
\ena
\ene
Then
\begin{equation} \label{tilde_phi_u vert}
 {\| \phivectilde_u \|}^2_{L^2(\Pi^a)}  \leq
\epsilon_3/H, 
\quad {\| \phivectilde_v \|}^2_{L^2(\Pi^a)} \leq
\epsilon_3/H, 
\end{equation}
where $ {\|  {\mathbf a}   \|}_{L^2(\Pi^a)}$ means $\lb   \int_{1/2}^1 dh
\int \int_D du dv\,  |{\mathbf a}|^2\rb^{\frac12}$.  
Since $u$ and $v$ are bounded on $D$, it follows that
\begin{equation}
\half{\|  \phivectilde_h \|}_{L^2 (\Pi^a)} \leq {\| h
\phivectilde_h \|}_{L^2(\Pi^a)}\leq (\epsilon_3/H)^{\frac12}
+ 
{\|u \phivectilde_u \|}_{L^2(\Pi^a)} +{\| v \phivectilde_v\|}_{L^2
  (\Pi^a)} \leq  
C^a_1
(\epsilon_3/H)^{\frac12} \label{tilde_phi_h vert}
\end{equation}
for some $C^a_1 >0$ independent of $\epsilon_3$, $H$ and
$\nvectilde$.

%\comment{$\mu,\xi,\eta\rightarrow u,v,h$}

$\phivectilde$ may be extended to a unit-vector field $\phivec$ on 
$\Lambda^a(H)$
according to
\begin{equation}
\phivec(u,v,h)= \phivectilde (u,v,s(h)), \quad 0 < h \leq 1,
\end{equation}
where $s(h)$ is a  $C^\infty$ function on $[0,1]$ with $\half \le s(h)
\le 1$, $s(0) = \half$, $s(h) = h$ for $\frac34\le h \le 1$, and with $s'(h)$ bounded away
from zero.
For  example, we can take 
\bge
\bga
s(h)= \frac{3}{4}+ \int_{\frac{3}{4}}^{h}  \al (t) dt,
\\
\al(x) := a+ (1-a) \frac{\int_{0}^{x}  f(t) dt}{\int_{0}^{\frac{3}{4}}  f(t) dt},
\\
f(x): = \lbr \bga \exp\ls \frac{1}{x-\frac{3}{4}}- \frac{1}{x-\frac{1}{2}}\rs, x\in (\frac{1}{2} , \frac{3}{4})
\\
0, \quad  x\not\in (\frac{1}{2} , \frac{3}{4}) \ena\rd,
\\
a := \frac{\ga-\frac{1}{4}}{\ga-\frac{3}{4}}, \ga := \frac{\int_{\frac{1}{2}}^{\frac{3}{4}}  f(t)(\frac{3}{4}-t) dt}{\int_{\frac{1}{2}}^{\frac{3}{4}}  f(t) dt} <\frac{1}{4}.\label{eq: s}
\ena
\ene
We define the corresponding extension of $\nvectilde$ by
\begin{equation}
  \label{eq:n(x,y,z)}
  \nvec(x ,y, z)= \phivec\left(\frac{x}{z},\frac{y}{z},\frac{z}{H}\right),
\quad 0 < z \leq H.
\end{equation}
Clearly, $\nvec$ is a smooth tangent unit-vector field on
$\Lambda^a(H)$. 
We estimate its energy (\ref{n_energyLa_small_vert}) as follows.
Let $h(s)$ be the inverse of $s(h).$ Since $h^2(s) s'(h(s))$ and
$1/s'(h(s))$ are bounded on $[0,1]$, say by $C_2$,
\begin{equation}\label{eq: E with elementary inequality}
\bga
E[ {\nvec}] = H \dst\int_{0}^1 dh \dst\int_D du dv  \lbr  \lb 
\phivec_u \rb^2 + \lb  \phivec_v \rb^2 + \lb  -u \phivec_u
-v \phivec_v + h \phivec_h\rb^2
\rbr 
=
\\
= H \dst\int_{\frac{1}{2}}^1  \frac{ds}{s^\prime (h(s))}\dst\int_{\mathit{D}} du dv  \lbr  \lb 
\phivectilde_u \rb^2 + \lb  \phivectilde_v \rb^2 + \lb  -u
\phivectilde_u -v \phivectilde_v +  h(s) s^\prime (h(s))
 \phivectilde_h \rb^2
\rbr \leq
\\
\leq  C_2 H \dst\int_{\frac{1}{2}}^1 dh \dst\int_{\mathit{D}}   du dv \lbr  \lb 
\phivectilde_u \rb^2 + \lb  \phivectilde_v \rb^2 + 3 \lb u
\phivectilde_u\rb^2 +3 \lb v \phivectilde_v \rb^2 +3  \lb 
 \phivectilde_h \rb^2
\rbr .
\ena
\end{equation}
In the last step of (\ref{eq: E with elementary inequality}), 
we have used the elementary inequality $(a + b +
c)^3 \le 3a^3 + 3b^3 + 3c^3$ (similar inqualities are used in what follows).
Using (\ref{tilde_phi_u vert}), \eqref{tilde_phi_h vert}
and the fact that $u$ and $v$ are
bounded on $D$, we obtain
\begin{equation}
  \label{eq:vertex result}
  E[ {\nvec}] \leq C_3^a H\lb {\| \phivectilde_u \|}^2_{L^2(\Pi^a)} + 
{\| \phivectilde_v \|}^2_{L^2(\Pi^a)}
+ {\| \phivectilde_h \|}^2_{L^2(\Pi^a)} \rb 
\leq C^a \epsilon_3,
\end{equation}
for some $C_3^a > 0$ and $C^a>0$ independent of $\epsilon_3$, $H$ and
$\nvectilde$.
\end{proof}

\begin{lem}\label{lem: edge}{\bf Edge extension.}  
  Let $E^b$ be an edge of $P$ between faces $F^+$ and $F^-$.  Let
  $(x,y,z)$ denote local Euclidean coordinates centred about the midpoint of
  $E^b$, with $z$-axis parallel to $E^b$ and positive $y$-axis
  directed into $P$ and lying in the midplane between $F^+$ and $F^-$,
  so that $F^{\pm}$ are given locally by $y \ge 0, x = \pm \tau y $, 
where $\tau = \tan\alpha$ and $2\alpha$ is the angle between $F^+$ and $F^-$.
  Let
  \begin{equation}
 \Lambda^b(W,L):=\lbr \label{eq: local edge coords}
  (x,y,z)\,\vert\, 0\leq y \leq W, -\tau y\leq x\leq \tau y,
-\half L \leq z\leq \half L
\rbr
\end{equation} 
denote the right prism whose axis has length $L$, is parallel to
$E^b$, and is symmetric about the midpoint of $E^b$, and whose
cross-section is an isosceles triangle with sides in $F^+$ and $F^-$ of
length $ W\sec \alpha$.  Let
\begin{equation}
  \label{eq:Pi^b}
  \Pi^b = \overline{
\Lambda^b(W,L) \setminus \Lambda^b(\half W,L)
}
\end{equation}
denote the closed interior half of this prism.
Let $\evec^b$ be a unit vector parallel to
  $E^b$.  Then, given a $C^\infty$ unit-vector field
$\nvectilde$
on $\Pi^b$
  which satisfies tangent boundary conditions on 
$F^\pm$, and constants $\epsilon_1 > 0$ and $0 <
  \epsilon_2 < \half$ such that
\begin{align}
E[\nvectilde]:=\dst\int_{\Pi^b} \lb \nabla \nvectilde\rb^2
&\leq\epsilon_1,
\label{n_energyPi_small} 
\\
\max_{\rvec\in \Pi^b} \vert \nvectilde(\rvec) - \evec^b\vert &\leq
\epsilon_2,
\label{n_jumpPi_small}
\end{align}
one can construct a $C^\infty$ unit-vector field $ \nvec$ in
$\Lambda^b(W,L)$  
satisfying tangent boundary conditions on $F^\pm$
and coinciding with 
$\nvectilde$ on $\Lambda^b(W,L)\setminus\Lambda^b(\smallfrac{3}{4} W,L)$ such that
\bge
E[{\nvec}]:=\dst\int_{\Lambda^b(W,L)} \lb \nabla \nvec\rb^2 \leq C^b 
(\epsilon_1 + L 
{\epsilon_2}^2),
\label{n_energyLa_small}
\ene
where $C^b$ is independent of $\nvectilde$, $\epsilon_1$,
$\epsilon_2$, $L$ and $W$.
\end{lem}
\begin{proof}
  It is convenient to introduce new  coordinates 
$(\xi,\eta,\mu)$ as follows:  
\begin{equation}
  \label{eq:edge coords}
  y =  W \eta, \quad  x =  \tau W \eta \xi, \quad z = L\mu .
\end{equation}
Let
\begin{equation}\label{eq: edge phitilde}
\phivectilde(\xi,\eta,\mu)= \nvectilde(\tau W\eta\xi , W\eta , L\mu).
\end{equation}
% that $\phi$ is supposed to be {\bf bold}, but it would not latex it
% that way..
Then 
\begin{equation}
 E[\nvectilde] 
= \int_{-\frac12}^{\frac12} d\mu \int_{-1}^1 d\xi \dst\int_{\frac{1}{2}}^1
\frac{d \eta}{\eta} 
\lbr
  \frac{L}{\tau}   
    (\phivectilde_\xi  )^2
  + L\tau
  (\eta \dst \phivectilde_\eta
     -\xi \dst\phivectilde_\xi
  )^2 
  + 
  \frac{W^2\tau}{L} \eta^2  ( \dst \phivectilde _\mu
  )^2 
\rbr 
\leq \epsilon_1.
\label{n_energyPi_newcoor}
\end{equation}
%% \begin{multline}
%%  E[\nvectilde] 
%% =\dst\int_{-\frac12}^{\frac12} d\mu \dst\int_{-1}^1 d\xi \dst\int_{\frac{1}{2}}^1
%% \frac{d \eta}{\eta} \times \\
%% \times 
%% \lbr
%%   \tau^{-1} L 
%%   \lb 
%%     \dst \partial_\xi \phivectilde  
%%   \rb^2
%%   + \tau^{-1} L 
%%   \lb 
%%      -\xi \dst\partial_\xi \phivectilde + \eta \dst\partial_\eta
%%      \phivectilde  
%%   \rb^2 
%%   + 
%%   (\tau W^2/L)\eta^2  \lb \dst \partial_\mu \phivectilde 
%%   \rb^2 
%% \rbr 
%% \leq \epsilon_1.
%% \label{n_energyPi_newcoor}
%% \end{multline}
%HERE
Let $ {\|  \avec   \|}_{L^2(\Pi^b)}$ denote the $L^2$-norm $\lb  \int_{-\frac12}^{\frac12} d\mu
\int_{-1}^1 d\xi \int_{\frac{1}{2}}^1 d\eta\, |\avec|^2   
\rb^{\frac{1}{2}}$ of a vector field $\avec$ on $\Pi^b$.  From
(\ref{n_energyPi_newcoor}) one can deduce that
\begin{equation}
{{\|   \phivectilde_\xi  \|}^2_{L^2(\Pi^b)}} \leq
C^b_1\frac{\epsilon_1}{L},\quad
{{\|  \phivectilde_\eta  \|}^2_{L^2(\Pi^b)}}  \leq
C^b_1\frac{\epsilon_1}{L},\quad
{{\| \phivectilde_\mu \|}^2_{L^2(\Pi^b)}}  \leq 
C^b_1 \frac{L\epsilon_1}{W^2}
\label{tilde_phi_xi_norm}
\end{equation}
for some $C^b_1 > 0$ independent of $\nvectilde$, $\epsilon_1$,
$\epsilon_2$, $L$ and $W$ (in fact, it suffices to take $C^b_1 =
4(\tau^\frac12 + \tau^{-\frac12})^2$).

We extend $\phivectilde$ to a tangent unit-vector field $\phivec$ on
$\Lambda^b(W,L)$ as follows.   In general, let $a_\parallel = \avec\cdot
\evec^b$ and $\avec_\perp = \avec - a_\parallel \evec^b$ denote the
components of $\avec$ parallel and perpendicular to $\evec^b$.
Let $\psi(\eta)$ be a $C^\infty$ map of the unit interval $[0,1]$ into
itself such that $\psi(\eta) =1$ for $0\leq\eta\leq \frac{1}{4}$ and
$\psi(\eta) = 0$ for $\frac{1}{2}\leq \eta\leq 1$. Let
$\Gamma(\xi,\eta,\mu) = (\xi, s(\eta), \mu)$, ie $\Gamma$ denotes the
change of coordinates $\eta \mapsto s(\eta)$ which leaves $\xi$ and
$\mu$ unchanged, where   $s(\eta)$ is a function like the one
described in \eqref{eq: s}.
Then $\phivec$ is
given by
\begin{equation} 
\phivec_\perp  = (1 - \psi)
\lb\phivectilde_\perp\circ \Gamma\rb,\quad
\phi_\parallel = \lb 1 - |\phivec_\perp|^2\rb^\frac12.
%\lb 1 -
%(1-\psi)^2 |\phivectilde_\perp\circ\Gamma|^2\rb^\frac12.
\label{eq:phitilde2}
\end{equation}
From \eqref{n_jumpPi_small} 
it follows that $|\phivectilde_\perp| \le
\epsilon_2 \le \half$ in $\Pi^b$, so that $\phi_\parallel$ is smooth.

The corresponding extenson
of $\tilde{\nvec}$ is given by
\begin{equation}
  \label{eq:ntilde2}
  \nvec(x,y,z) = \phivec\lb \frac{x}{\tau y},\frac{y}{W},
  \frac{z}{L}\rb.
\end{equation}
In analogy with \eqref{n_energyPi_newcoor}, 
the energy of $\nvec$ is given by
\begin{equation}
E[ \nvec ] = \int_{-\frac12}^{\frac12} d\mu \int_{-1}^1 d\xi
\int_{0}^{1} 
\frac{d \eta}{\eta}
\lbr  \frac{L}{\tau} \lb \phivec_\xi \rb^2 +
L\tau \lb 
\eta \phivec_\eta - \xi   \phivec_\xi \rb^2 +  
\frac{W^2\tau}{L}
\eta^2
\lb
\phivec_\mu \rb^2 \rbr.
\label{n_energyLa_newcoor}
\end{equation}
Noting that $\phivec = \evec^b$ for $\eta\le \fourth $ (so that
integrand vanishes in this range), 
%and making use
%of the elementary inequality $|\avec + \bvec|^2 \le 2|\avec|^2 + 2|\bvec|^2$, 
one can obtain the estimate 
\begin{equation}
  \label{eq:E_estimate_2}
E[ \nvec ] \le C^b_2 \int_{-\frac12}^{\frac12} d\mu \int_{-1}^1 d\xi
\int_{0}^{1} 
d \eta
\lbr  L \lb \phivec_\xi \rb^2 +
L \lb \phivec_\eta\rb ^2 + 
\frac{W^2}{L}
\lb
\phivec_\mu \rb^2 \rbr
\end{equation}
for some $C^b_2 > 0$ independent of $\nvectilde$, $\epsilon_1$,
$\epsilon_2$, $L$ and $W$ (in fact, it suffices to take $C^b_2 = 8\tau + 4\tau^{-1}$).

The derivatives of $\phivec$ can be expressed in terms of derivatives
of $\phivectilde$ as follows.  We have that $\phivec_{\perp\xi}
= (1-\psi)(\phivectilde_{\perp\xi}\circ\Gamma)$, while 
$\phi_{\parallel\xi} =
(1-\psi)^2((\phitilde_\parallel\, \phitilde_{\parallel\xi})\circ\Gamma)/\phi_\parallel$,
where we have used $|\phivectilde_\perp|^2 + {\phitilde_\parallel}^2 =
1$.  Since $|1-\psi|^2 \le 1$ and $|(\phitilde_\parallel\circ
\Gamma)/\phi_\parallel| \le 1$, it follows that
\begin{equation}
  \label{eq:partial_xi bound}
  (\phivec_\xi)^2 = (\phivec_{\perp\xi})^2 + (\phi_{\parallel\xi})^2 
\le (\phivectilde_\xi\circ\Gamma)^2.
\end{equation}
A similar argument shows that
\begin{equation}
  \label{eq:partial_mu bound}
  ( \phivec_\mu)^2 \le (\phivectilde_\mu\circ\Gamma)^2.
\end{equation}
Next, we have that
\begin{equation}
  \label{eq:partial_eta perp}
  \phivec_{\perp\eta} = -\psi'(\phivectilde_\perp\circ\Gamma) +
  (1-\psi)\,s'\,
(\phivectilde_{\perp\eta}\circ\Gamma),
\end{equation}
which implies that
\begin{equation}
  \label{eq:partial_eta perp est}
  (\phivec_{\perp\eta})^2 \le
  2{\psi'}^2|\phivectilde_\perp\circ\Gamma|^2 + 2{s'}^2(\phivectilde_{\perp\eta}\circ\Gamma)^2.
\end{equation}
Similarly,  
\begin{equation}
  \label{eq:partial_eta para}
  \phi_{\parallel\eta} =
  (1-\psi)\psi'\lb|\phivectilde_\perp|^2
\circ\Gamma\rb/\phi_\parallel +
  (1-\psi)^2\,s'\,
((\phitilde_\parallel\,\phitilde_{\parallel\eta})\circ\Gamma)/\phi_\parallel.
\end{equation}
From \eqref{n_jumpPi_small}, $|\phivectilde_\perp|^2\le {\epsilon_2}^2
< \fourth$ and $|\phi_\parallel| >\half$, so that
\begin{equation}
  \label{eq:partial_eta para est}
  (\phi_{\parallel\eta})^2 \le
  2{\psi'}^2{\epsilon_2}^2 + 2{s'}^2(\phitilde_{\parallel\eta}\circ\Gamma)^2.
\end{equation}
Together, \eqref{eq:partial_eta perp est} and \eqref{eq:partial_eta
  para est} give
\begin{equation}
  \label{eq:partial_eta est}
  (\phivec_\eta )^2 \le 4 C_3 {\epsilon_2}^2 +
  2{s'}^2(\phivectilde_\eta\circ \Gamma)^2,
\end{equation}
where $C_3$ bounds ${\psi'}^2$.

We substitute \eqref{eq:partial_xi bound}, \eqref{eq:partial_mu bound}
and \eqref{eq:partial_eta est} into the estimate
\eqref{eq:E_estimate_2}.  Replacing $\int_0^1 d\eta$ by
$\int_{\frac12}^1\eta'(s)ds$, we get 
\begin{equation}
  \label{eq:E est penultimate}
  E[\nvec] \le C^b_2 C_4 
\lb
L {\|  \phivectilde_\xi \|}^2_{L^2(\Pi^b)} +
L \lb  4C_3 {\epsilon_2}^2 + 2{\|\phivectilde_\eta
  \|}^2_{L^2(\Pi^b)}\rb + 
\frac{W^2}{L} {\|  \phivectilde_\mu \|}^2_{L^2(\Pi^b)}
\rb,
\end{equation}
where $C_4$ bounds both $s'$ and $\eta' = 1/s'$.  From
\eqref{tilde_phi_xi_norm},
\begin{equation}
  \label{eq:last line in lemma2}
  %E[\nvec] \le 4C^b_1 C^b_2 C_4 (\epsilon_1 + LC_3{\epsilon_2}^2)
      E[\nvec] \le 4C^b_1 C^b_2 C_4 \epsilon_1 + 4LC_2^bC_3C_4{\epsilon_2}^2,
\end{equation}
which implies the required result.
\end{proof}

\begin{thm}\label{thm: thm 2}
Suppose the tangent unit-vector field $\nvectilde$
is continuous on  $P$ with
$ {\| \nvectilde\|}^2_{1,2}(P) =   \int_P \lb\nabla
\nvectilde\rb^2 <\infty.$ Then for all  $\epsilon > 0$, there exists a
$C^\infty(P)$ 
 tangent unit-vector field $\nvec$  homotopic to
$\nvectilde$ with
$\|\nvectilde-\nvec\|_{1,2}(P) \leq \epsilon .$
\end{thm}

\begin{rem} In general, $\nvec(\rvec) -
\nvectilde(\rvec)$ is not uniformly small in $\rvec$; in small
neighbourhoods of the edges and vertices, this difference may be of
order one.  However, the contribution of these neighbourhoods to the
Sobolev norm, $\|\nvectilde-\nvec\|_{1,2}(P)$, is small.
\end{rem}
\begin{proof} 
  {\em Step 0: truncation, reflection}.  Let $\Lambda^a(H)$ denote the
  vertex prisms defined in Lemma~\ref{lem: vertex extension}.  We 
  choose $H$ sufficiently small so that these  do not intersect, and so
  that, for given $\epsilon_1 > 0$, 
\begin{equation}\label{eq: nvectilde est on Lambda^a}
  \| \nvectilde\|^2_{1,2}\lb\cup_a\Lambda^a(H)\rb \leq  \epsilon_1.
\end{equation}
%This is possible since $ \| \nvectilde\|_{1,2}$ is finite.
%% In fact, don't need to introduce $\delta_c$ yet.
%% Let
%% \begin{equation}
%%   \label{eq:P'}
%%   P' = \overline{ {P/\lb \cup_{a}\Lambda^{a}(H) \rb}}
%% \end{equation}
%% denote the closed polyhedron obtained by removing the 
%% $\Lambda^{a}(H)$'s from $P$.  Since $\nvectilde$ is continuous
%% on $P,$ it is uniformly continuous on $P'$. Choose $\delta_c > 0$ 
%% such that 
%% \bge 
%% \rvec, \svec \in P', |\rvec - \svec| \le
%% \delta_c \implies
%% \lv \nvectilde(\rvec) -
%% \nvectilde(\svec)\rv \leq \half \eps_2 < \fourth.  
%% \ene

Let $\Lambda^b(W,\Ltilde^b)$ denote the edge prisms defined in
Lemma~\ref{lem: edge}.  We take these to have the same width, $W$, but
allow them to have different lengths $\Ltilde^b$.  We choose $W$
sufficiently small and the $\Ltilde^b$'s so that the following
conditions are satisfied: First, the $\Lambda^b(W,\Ltilde^b)$'s do not
interesect.  Second, the top and bottom faces of
$\Lambda^b(W,\Ltilde^b)$ (given by $z = \pm \half\Ltilde^b$ in the
local Euclidean coordinates of (\ref{eq: local edge coords})) are
contained in the respective vertex prisms
$\Lambda^a(\smallfrac{3}{12}H)$ and $\Lambda^{a'}(\smallfrac{3}{12}H)$
at the endpoints of $E^b$, but do not intersect the smaller vertex
prisms $\Lambda^a(\smallfrac{2}{12}H)$ and
$\Lambda^{a'}(\smallfrac{2}{12}H)$.  
Third, for given $\epsilon_1 > 0$, 
\begin{equation}
  \label{eq:condition 3}
  \| \nvectilde\|_{1,2}^2\lb\cup_b\Lambda^{b}(W,\Ltilde^b)\rb \leq
\half\epsilon_1.
\end{equation}
Fourth, for given $0 < \epsilon_2 < \half$,
\begin{equation}
  \label{eq:condition 4}
  \max_{\rvec\in \Lambda^b(W,\Ltilde^b)} \lv {\nvectilde}(\rvec)
  - \evec^b\rv \le \half \epsilon_2 < \fourth,
\end{equation}
where $\evec^b$ is
  the edge orientation of ${\nvectilde}$ on $E^b$.

Let 
\begin{equation}
  \label{eq:Ptilde}
  P'= \overline{P
\setminus\lb \cup_a
\Lambda^{a}(\smallfrac{4}{12} H)\cup_b \Lambda^{b}(\half W, \Ltilde^b)\rb}
\end{equation}
be the closed polyhedron obtained by removing vertex and edge prisms
of indicated size from $P$.  Choose $\Delta_1 > 0$ sufficiently small
so that, for all $\rvec \in P'$, $B(\rvec,\Delta_1)$ -- the
$\Delta_1$-ball centered at $\rvec$ -- intersects at most one face
of $P$.  Let 
\begin{equation}
  \label{eq:P' + delta_1}
  P' + \Delta_1 = \lbr \rvec\in \Rr^3 \big |\, |\rvec - \rvecp| \le \Delta_1
  \ \text{for some}\ \rvecp \in P' \rbr
\end{equation}
denote the closed $\Delta_1$-neighbourhood of
$P'$.  We define a unit-vector field $\nvectilde_+$ on 
$P' + \Delta_1$ as follows.  If $\rvec \in P$,
$\nvectilde_+(\rvec)$ is just taken to be $\nvectilde(\rvec)$.  If
$\rvec \in (P' + \Delta_1)\setminus P$, then we must have that 
$\rvec = \pvec + \alpha \Fvec^c$ for some (uniquely determined) $\pvec$
on a face $F^c$ of $P$
and for some $0 <\alpha \le \Delta_1$.  In this case, we take
$\nvectilde_+(\pvec + \alpha \Fvec^c) = {\cal R}\,\cdot\, \nvectilde_+(\pvec -
\alpha \Fvec^c)$, where ${\cal R}$ denotes reflection about the plane
normal to
$\Fvec^c$.  
It is clear
that the Sobolev norm $\| \nvectilde_+
\|_{1,2} (P' + \Delta_1)$ of the extended vector field is finite.
Tangent boundary conditions imply that
$\nvectilde_+$ is continuous at points $\pvec\in P'$ 
that lie on a  face of $P$.
Moreover, 
the average of $\nvectilde_+$ over 
$B(\pvec,\delta)$, where $\delta < \Delta_1$,
is tangent
to the face.  This remains true
for weighted averages over $B(\pvec,\delta)$ provided the weights at $\pvec
+ \alpha \Fvec^c$ and $\pvec
- \alpha \Fvec^c$ are equal.\\

\noindent {\em Step 1: Bulk.} 
Since $\nvectilde$ is continuous
on $P$, $\nvectilde_+$ is continuous on $P'+\Delta_1$, and
therefore uniformly continuous (since $P'+\Delta_1$ is compact).
Choose $\Delta_2 >
0$ sufficiently small so that 
 \begin{equation} 
 \rvec, \rvecp \in P'+ \Delta_1, \quad |\rvec - \rvecp| \le
 \Delta_2 \implies
 \lv \nvectilde_+(\rvec) -
 \nvectilde_+(\rvecp)\rv \leq \fourth \epsilon_2.
 \end{equation}
Let $K(r)\geq 0$ be a smooth function with support contained in
$(0,1)$, normalized so that
\begin{equation}
  \label{eq:K(r)}
  4\pi \int_0^\infty K (r) r^2 dr =1.
\end{equation}
Choose $\delta>0$ so that $\delta < \min(\Delta_1,\Delta_2).$
We construct a smooth unit-vector field $\uvec$  on $P'$ by averaging
$\nvectilde_+$ over balls of radius $\delta$, as follows:
\bge
\uvec(\rvec) = 
\int_{\Rr^3}
\frac{1}{\delta^3} K\lb \frac{\lv \rvec - \rvecp \rv}{\delta}\rb
\nvectilde_{+}(\rvecp) \, dV'.\label{eq:uvec}
\ene
Uniform continuity ensures that $|\uvec(\rvec) - \nvectilde(\rvec)| \le
\fourth\epsilon_2< \smallfrac{1}{8}$ for  $\rvec \in P'$, and in particular that
$\uvec \ne 0$.  Therefore, the unit-vector field
\begin{equation}
  \label{eq:vvec}
  \vvec := \uvec/|\uvec|
\end{equation}
is smooth, with  
\begin{equation}
  \label{eq:vvec - ntilde on P'}
 \max_{\rvec\in P'} | \vvec(\rvec) - \nvectilde(\rvec)| \le
\half\epsilon_2.
\end{equation}
$\vvec$ satisfies tangent boundary conditions at points $\pvec$ in $P'$
on faces $F^c$ of $P$ (since $K(\alpha \Fvec^c/\delta) = 
K(-\alpha \Fvec^c/\delta)$).

It is straightforward to show (the arguments are similar to those of the
Meyers-Serrin theorem \cite{adams}, see also Appendix A, \cite{bl})
that 
$\|\vvec - \nvectilde \|_{1,2} (P')$ 
can be made arbitrarily small
with $\delta$.  Indeed, from \eqref{eq:uvec}, for $\rvec \in P' $,
\begin{multline}
  \label{eq:pointwise grad difference}
  \nabla \uvec(\rvec) - 
\nabla  \nvectilde(\rvec) = 
\int_{\Rr^3}
\frac{1}{\delta^3}K\lb \frac{\lv \rvec - \rvecp \rv}{\delta}\rb
\lb
\nabla\nvectilde_{+}(\rvecp) -
\nabla\nvectilde_{+}(\rvec)\rb
\, d^3 {r'} = \\
= \int_{\Rr^3}
\ls 
\frac{1}{\delta^{3/2}} 
K^\half\lb \frac{\lv \rvec - \rvecp \rv}{\delta}\rb \rs
\ls
\frac{1}{\delta^{3/2}} 
K^{\smallfrac{1}{2}}\lb \frac{\lv \rvec - \rvec' \rv}{\delta}\rb
\lb \nabla\nvectilde_{+}(\rvecp) -
\nabla\nvectilde_{+}(\rvec)\rb
\rs
\, d^3 {r'}.
\end{multline}
The Cauchy-Schwartz inequality gives that 
\begin{equation}
  \label{eq:last in line}
|\nabla \uvec(\rvec) - 
\nabla \nvectilde(\rvec)|^2 \le \int_{\Rr^3}\frac{1}{\delta^3} 
K\lb \frac{\lv \rvec - \rvecp \rv}{\delta}\rb
\lv
\nabla \nvectilde_+(\rvec) -
\nabla \nvectilde_+(\rvecp)
\rv^2 \, d^3 r'.
\end{equation}
 Since  square-integrable vector fields can be
approximated arbitrarily closely (with respect to the $L^2$-norm) by
continuous vector fields, we may write $\nabla \nvectilde_+ ={\mathbf c} + {\mathbf h}, 
$ where ${\mathbf c}$ is continuous, and therefore uniformly continuous, on $P' +
\Delta_1$, and $\| {\mathbf h}\|_{L^2}(P'+\Delta_1)$
is arbitrarily small. Extend 
${\mathbf h}$ by zero outside $P'+\Delta_1.$ Then
\begin{multline}
\|\nabla \uvec - 
\nabla \nvectilde_+\|^2_{L^2}(P')\leq 
\int_{P'}  d^3 r \int_{\Rr^3}  d^3r' \frac{1}{\delta^3} 
K\lb \frac{\lv \rvecp - \rvec \rv}{\delta}\rb
\lv
\nabla \nvectilde_+(\rvec) -
\nabla \nvectilde_+(\rvecp)
\rv^2 \leq \notag \\
\leq  \int_{P'}  d^3 r \int_{\Rr^3} d^3 r'  \frac{1}{\delta^3} 
K\lb \frac{\lv \rvec - \rvecp \rv}{\delta}\rb
3\lb  |{\mathbf h}(\rvec)|^2 + |{\mathbf h} (\rvecp)|^2 + \lv 
 {\mathbf c}(\rvec) -
 {\mathbf c}(\rvecp)
\rv^2 
\rb \leq
\notag \\
\leq
 3 \| {\mathbf h}\|^2_{L^2}(P'+\Delta_1) + 3 \| {\mathbf
  h}\|^2_{L^2}(P') +
3 \int_{P'} d^3 r\int_{\Rr^3}d^3 r' \frac{1}{\delta^3} 
K\lb \frac{\lv \rvec - \rvecp \rv}{\delta}\rb
\lv
 {\mathbf c}(\rvec) -
 {\mathbf c}(\rvecp)
\rv^2. 
\end{multline}

The first two terms can be made arbitrarily small, and since ${\mathbf c}$ is uniformly
continuous, the last term approaches zero with $\delta$.  That the
same is true for $\vvec = \uvec/|\uvec|$ is easily established; note
that
$\nabla \lv
\uvec \rv^2 = \nabla \lb(\uvec-
\nvectilde_+)\cdot(\uvec + \nvectilde_+)\rb$,
and  that $(\uvec + \nvectilde_+)$ is uniformly
bounded on $P'$ while $( \uvec-
\nvectilde_+)$ approaches zero uniformly with $\delta$.
Thus, we can choose
$\delta$ small enough so that
\begin{equation}
  \label{eq:vvec-ntilde}
  \|\vvec -
\nvectilde \|_{1,2}^2 (P')\le \half\epsilon_1.
\end{equation}
\\
\noindent {\em Step 2: Edges.} Let $\Lambda^{b}(W,L^b)$ be  edge
prisms shorter than those introduced in  Step 0, ie $L^b <
\Ltilde^b$, such that their top and bottom faces are contained in
$\cup_a \Lambda^a(\smallfrac{5}{12}H)\setminus\Lambda^a(\smallfrac{4}{12}H)$.
Let $\Pi^b = \overline{\Lambda^b(W,L^b)\setminus\Lambda^b(\frac12 W,L^b)}$ denote the
half-prism, as in Lemma~\ref{lem: edge}, and let
\begin{equation}
  \label{eq:Lambda and Pi}
  \Lambda_E = \cup_b \Lambda^b(W,L^b), \quad \Pi_E = \cup_b \Pi^b
\end{equation}
denote the union of the edge prisms and half-prisms respectively. 
 $\vvec$ defines a  smooth tangent unit-vector field on
$\Pi_E$.
From \eqref{eq:condition 3} and \eqref{eq:vvec-ntilde}, its energy is
bounded by
\begin{equation}
\|\vvec\|_{1,2}^2(\Pi_E)  
\leq 
2\|\vvec - \nvectilde  \|^2_{1,2}(\Pi_E) +
2\| \nvectilde  \|^2_{1,2} (\Pi_E) \leq
\epsilon_1+  \epsilon_1 \leq 2\epsilon_1.
\end{equation}
From \eqref{eq:condition 4} and \eqref{eq:vvec - ntilde on P'}, 
\begin{equation}
  \label{eq:second condition of lemma 2}
\max_{\rvec \in \Pi^b}
\left| 
\vvec(\rvec) - 
\evec^b \right| \le 
\max_{\rvec \in \Pi^b}
\lb
\left| 
\vvec(\rvec) - 
\nvectilde(\rvec) \right| +
\left| 
\nvectilde(\rvec) - 
\evec^b \right| 
%\le 
\rb
\leq \half \epsilon_2+ \half \epsilon_2 = \epsilon_2.
\end{equation}
Thus, the conditions of Lemma~\ref{lem: edge} are  satisfied for each
$\Lambda^b(W,L^b)$, and 
we can construct a smooth tangent unit-vector field $\wvec$ on 
$\Lambda_E$ which coincides with $\vvec$ on
$\Lambda_E\setminus(\cup_b \Lambda^b(\smallfrac{3}{4} W, L^b))$
such that
\begin{equation}
\label{eq: wvec small}
\|\wvec \|_{1,2}^2 \lb \Lambda_E  \rb \leq 
{\textstyle \sum_b} C^b ( 2 {\epsilon_1} + L^b{\epsilon_2}^2),
\end{equation}
where $C^b > 0$ is independent of $\epsilon_1$, $\epsilon_2$,
$\nvectilde$, $W$ and $L^b$.
We may extend $\wvec$ (smoothly) to  
$  P'' = \overline{
P\setminus\lb \cup_a \Lambda^a(\smallfrac{5}{12} H)\rb}$
by taking $\wvec$ to coincide with
$\vvec$ on  $P''\setminus\Lambda_E$. 
From \eqref{eq:condition 3},
\eqref{eq:vvec-ntilde} 
and \eqref{eq: wvec small},
\begin{align} \label{eq: wvec - nvectilde estimate}
\|\wvec -\nvectilde \|_{1,2}^2 \lb P'' \rb &=
 \|\wvec -\nvectilde \|_{1,2}^2 \lb P''\setminus \Lambda_E \rb 
+  \|\wvec -\nvectilde \|_{1,2}^2 \lb \Lambda_E \rb \nonumber
\\
&\leq \|\vvec -\nvectilde \|_{1,2}^2 \lb P''\setminus \Lambda_E \rb 
+ 2\lb \|\wvec\|_{1,2}^2 + \|\nvectilde\|_{1,2}^2\rb \lb \Lambda_E \rb \notag
\\
&\leq \lb \smallfrac{3}{2} + 4{\textstyle \sum_b} C^b\rb \epsilon_1 + 2 {\textstyle \sum_b}
C^b L^b{\epsilon_2}^2.
%\rb
\end{align}
In the second inequality we have used the fact that $P''\setminus
\Lambda_E \subset P'$.

{\em Step 3. Vertices.} 
Let $\Pi^a = \overline{\Lambda^a(H)\setminus \Lambda^a(\frac12 H)}$ denote the half-prism,
as in Lemma~\ref{lem: vertex extension}, and let
\begin{equation}
  \label{eq:Lambda and Pi vertices}
  \Lambda_V = \cup_a \Lambda^a(H), \quad \Pi_V = \cup_a \Pi^a
\end{equation}
denote the union of the vertex prisms and half-prisms respectively.
From \eqref{eq: nvectilde est on Lambda^a} and 
\eqref{eq: wvec - nvectilde estimate},
\begin{multline}
\| \wvec \|^2_{1,2} \lb \Pi_V \rb 
\leq 2 \|\wvec -\nvectilde \|^2_{1,2} \lb \Pi_V \rb + 
2 \|\nvectilde \|^2_{1,2} \lb \Pi_V \rb  \leq \\ \leq
\lb \lb 5  + 8{\textstyle \sum_b} C^b \rb \epsilon_1 + 
4{\textstyle \sum_b}
C^b L^b{\epsilon_2}^2\rb
:= \epsilon_3.
\end{multline}
Thus, the conditions of Lemma~\ref{lem: vertex extension} are satisfied for
each $\Lambda^a(H)$, and 
we can construct a smooth tangent unit-vector
field $\nvec$ on $\Lambda_V$ coinciding with $\wvec$ on 
$\Lambda_V\setminus(\cup_a\Lambda^a(\smallfrac{3}{4} H)$ such that
\begin{equation} \label{eq: nvec est }
\|\nvec \|^2_{1,2} \lb\Lambda_V \rb \leq {\textstyle \sum_a} C^a
{\epsilon_3}
\end{equation}
where $C^a > 0$ is independent of $H$, $\epsilon_3$ and $\nvectilde$.

We extend $\nvec$ (smoothly) to all of $P$ by taking $\nvec$ to
coincide with $\wvec$ on $P\setminus\Lambda_V$.  From \eqref{eq: nvectilde est on Lambda^a},
\eqref{eq: wvec -
  nvectilde estimate} and \eqref{eq: nvec est }, 
\begin{align} 
%\label{eq: nvec - nvectilde estimate}
\|\nvec -\nvectilde \|_{1,2}^2 
\lb P \rb 
& =   \| \nvec -\nvectilde \|_{1,2}^2 
\lb P\setminus \Lambda_V \rb 
+  \|\nvec -\nvectilde \|_{1,2}^2 
\lb \Lambda_V \rb \notag
\\
& \leq   \|\wvec -\nvectilde \|_{1,2}^2 
\lb P\setminus \Lambda_V \rb 
+ 2
\lb \|\nvec\|_{1,2}^2 + \|\nvectilde\|_{1,2}^2\rb 
\lb \Lambda_V \rb \notag
\\
&\leq 
\lb \half + 2{\textstyle \sum_a} C^a \rb 
\epsilon_3 +  \epsilon_1.
\end{align}
By choosing $\epsilon_1$ and $\epsilon_2$ appropriately, 
$\|\nvec -\nvectilde \|_{1,2}^2$ can be made arbitrarily small.
From \eqref{eq:vvec - ntilde on P'}, $\lv \nvec -\nvectilde \rv \leq \half \epsilon_2$
on $P'$.  This ensures that $\nvec$ and $\nvectilde$ have the
same kink numbers and wrapping numbers.  The construction also ensures
that their edge orientations are the same.
\end{proof}

\section{Discussion}\label{sec:discussion}

We have derived a lower bound, depending on homotopy type, for the
energies of harmonic maps of convex polyhedra in $\Rr^3$ to the unit
sphere $S^2$ which satisfy tangent boundary conditions and are
continuous everywhere except at vertices.  
It is natural to ask whether this lower bound is sharp.  For the case
of a rectangular prism, eg a cube, numerical calculations and
analytical arguments have indicated that, for a large set of homotopy
types, the lower bound (\ref{low_bound}) is not sharp, but differs
from the actual lower bound by a fixed quantity independent of
homotopy type \cite{mrz}.  A related question is whether the infimum
(\ref{energyMinimimal}) is achieved by some $\nvec \in \czero(P)\cap
W^{1,2}(P)$.  Numerics and analytic arguments for a rectangular prism
have suggested that for a small number of ``unwrapped'' states, the
infimum is achieved, but for a large set of homotopy types, a
sequence of configurations with energies approaching $M(h)$ develop
discontinuities along the edges.  We plan to present these results in
a subsequent paper.

While our present motivation comes from liquid crystal physics,
related considerations for maps of two- (three-) dimensional polyhedra
to spheres may be relevant for string (M-) theories with tangent
boundary conditions (`$T$ branes').

\section*{Acknowledgment}\label{sec:acknowledgment}
We are grateful to Adrian Geisow and Chris Newton, Hewlett-Packard
Laboratories, Bristol for stimulating out interest in this area. 
AM was supported by an
Industrial CASE studentship funded by EPSRC and by Hewlett-Packard
Laboratories, Bristol.


\begin{thebibliography}{5}
\bibitem{dg} 
C Oseen, Theory of Liquid Crystals, {\it Trans. Faraday Soc} {\bf 29}, 883-899, 1933.
\\
F.C. Frank, On the theory of liquid crystals, {\it Discuss. Faraday Soc.} {\bf 25},19-28, 1958.
\\
L.D. Landau and  E.M. Lifshitz, {\it Theory of elasticity},
  3rd English ed., Pergamon Press, 1986.
  \\
  P.G. de Gennes and J. Prost, {\it The physics of liquid crystals}, 2nd ed,  Oxford, Clarendon
  Press, 1995.  

\bibitem{rz} J.M. Robbins and M. Zyskin,
  Classification of unit-vector fields in convex polyhedra with
  tangent boundary conditions,  Preprint math-ph/0402025.

\bibitem{newtonspiller} C.J.P. Newton and T.P. Spiller,
  Proc. 17th IDRC (SID), 1997.

\bibitem{mrz} A. Majumdar, J.M. Robbins and M. Zyskin, in preparation.
\bibitem{bl} H. Brezis, J.M. Coron and E. Lieb, Harmonic maps with
  defects, 
{\it Comm. Math. Phys.}  {\bf 107} 649-705, 1986.
\bibitem{adams} R.A. Adams, {\it Sobolev spaces}, Academic Press, 1975. 
\end{thebibliography}
\end{document}